\documentclass[aps,prl,twocolumn,amssymb,showpacs,superscriptaddress]{revtex4}

\usepackage{graphicx}
\usepackage[english]{babel}
\usepackage{bm}    

\def\reff#1{(\ref{#1})}
\newcommand{\be}{\begin{equation}}
\newcommand{\ee}{\end{equation}}
\newcommand{\<}{\langle}
\renewcommand{\>}{\rangle}

\def\spose#1{\hbox to 0pt{#1\hss}}
\def\ltapprox{\mathrel{\spose{\lower 3pt\hbox{$\mathchar"218$}}
 \raise 2.0pt\hbox{$\mathchar"13C$}}}
\def\gtapprox{\mathrel{\spose{\lower 3pt\hbox{$\mathchar"218$}}
 \raise 2.0pt\hbox{$\mathchar"13E$}}}

\newcommand{\scrd}{{\cal D}}

\newcommand{\scrm}{{\cal M}}
\newcommand{\scrn}{{\cal N}}

\newcommand{\scrs}{{\cal S}}

\def\bsigma{{\boldsymbol{\sigma}}}

\def\R{{\mathbb R}}

\begin{document}

\title{Cluster simulations of loop models on two-dimensional lattices}

\author{Youjin Deng}
\affiliation{Department of Physics, New York University,
      4 Washington Place, New York, NY 10003, USA}
\author{Timothy M.~Garoni}
\affiliation{Department of Physics, New York University,
      4 Washington Place, New York, NY 10003, USA}
\author{Wenan Guo}
\affiliation{Physics Department, Beijing Normal University,
      Beijing 100875, China}
\author{Henk W.J. Bl\"ote}
\affiliation{Faculty of Applied Sciences, Delft University of Technology,
    P.O.~Box 5046, 2600 GA Delft, The Netherlands}
\affiliation{Lorentz Institute, Leiden University,
    P.O.~Box 9506, 2300 RA Leiden, The Netherlands}
\author{Alan D. Sokal}
\affiliation{Department of Physics, New York University,
      4 Washington Place, New York, NY 10003, USA}
\affiliation{Department of Mathematics,
      University College London, London WC1E 6BT, UK}

\date{November 20, 2006, revised January 16, 2007}

\begin{abstract}
We develop cluster algorithms for a broad class of
loop models on two-dimensional lattices,
including several standard $O(n)$ loop models at $n \ge 1$.
We show that our algorithm has little or no critical slowing-down
when $1 \le n \le 2$.
We use this algorithm to investigate the honeycomb-lattice $O(n)$ loop model,
for which we determine several new critical exponents,
and a square-lattice $O(n)$ loop model,
for which we obtain new information on the phase diagram.
\end{abstract}

\pacs{05.10.Ln, 05.50.+q, 64.60.Cn, 64.60.Fr, 75.10.Hk}

\maketitle

{}From the beginning of the theory of critical phenomena,
two models have played a central role:
the $q$-state Potts model \cite{Potts_52,Wu_82+84}
and the $O(n)$ spin model \cite{Stanley_68,Pelissetto-Vicari}.
The parameter $q$ or $n$ is initially a positive integer,
but the Fortuin--Kasteleyn (FK) representation \cite{FK_69+72}
and the loop representation \cite{Domany_81}
show, respectively, how the models can be extended to
arbitrary real or even complex values of $q$ and $n$ \cite{note_loop_rep}.
In particular, for $q,n > 0$ the extended model has a probabilistic
interpretation as a model of random geometric objects:
clusters \cite{Grimmett_06} or loops \cite{Nienhuis_82+84}, respectively.
These geometric models play a major role in recent developments of
conformal field theory \cite{DiFrancesco_97}
via their connection with stochastic Loewner evolution (SLE)
\cite{SLE_math,SLE_phys}.

Since nontrivial models of statistical mechanics are rarely exactly soluble,
Monte Carlo simulations have been an important tool for obtaining information
on phase diagrams and critical exponents \cite{Binder_79-92}.
Unfortunately, Monte Carlo simulations typically suffer from severe
{\em critical slowing-down}\/, so that the computational efficiency
tends rapidly to zero as the critical point is approached \cite{Sokal_Cargese}.
An important advance was made in 1987 with the invention of
the Swendsen--Wang (SW) algorithm \cite{Swendsen_87}
for simulating the ferromagnetic Potts model at positive integer $q$,
based on passing back and forth between the spin and FK representations.
The SW algorithm does not eliminate critical slowing-down,
but it radically reduces it \cite{Ossola-Sokal}.
Since then, many similar ``cluster algorithms'' have been devised,
based on this principle \cite{auxiliary}
of augmenting the original spin model with auxiliary variables
and then passing back and forth.
But cluster algorithms have traditionally been limited to integer $q$,
since they make essential use of the spin representation.

This limitation was first overcome in 1998
by Chayes and Machta \cite{Chayes-Machta},
who devised a cluster algorithm for simulating
the FK random-cluster model at any real $q \ge 1$.
For loop models, by contrast, efficient simulation at noninteger $n$
has remained out of reach;
to our knowledge only two Monte Carlo simulations at $n \neq 1$
have ever been published \cite{Karowski_83,Ding_06},
and they used local algorithms.
(Instead, numerical transfer-matrix techniques have been employed
 \cite{TM_loop_studies}.)
As a result, many open questions remain:
for instance, the nature of the phase transition is unclear for the
$n > 3/2$ honeycomb-lattice loop model with vacancies \cite{Guo_06};
and the phase diagrams and universality classes of loop models
on lattices other than honeycomb are largely unexplored \cite{Chayes_00}.

In this Letter we shall set forth a broad (but non-specific) generalization
of the Chayes--Machta idea,
and then show how it can be adapted to provide a cluster algorithm
for simulating loop models on two-dimensional lattices at any real $n \ge 1$.
We shall furthermore present numerical evidence that
this cluster algorithm has little or no critical slowing-down
when $1 \le n \le 2$.
Finally, we shall use this algorithm to obtain new results
for the phase diagram
of a loop model on the square lattice.

We begin by considering a {\em generalized random-cluster (RC) model}\/,
defined as follows:
Let $G=(V,E)$ be a finite graph with vertex set $V$ and edge set $E$.
A configuration of the model is specified by a subset $A \subseteq E$
(the set of ``occupied bonds''), and the partition function is
\be
   Z  \;=\;
   \sum_{A \subseteq E}
       \left( \prod_{e \in A} v_e \!\right)
       \!
       \left( \prod_{i=1}^k W(H_i) \!\right)
       \,,
 \label{def.genRC}
\ee
where $H_1,\ldots,H_k$ are the connected components of the graph $(V,A)$;
here $\{ v_e \}$ are nonnegative edge weights,
and $\{ W(H) \}$ are nonnegative weights associated to the connected subgraphs
$H$ of $G$.
The model \reff{def.genRC} reduces to the FK random-cluster model
if $W(H) = q$ for all $H$;
other special cases include an FK representation
for the Potts model in a magnetic field,
$W(H) = q-1 + e^{h |V(H)|}$ \cite{Sokal_chromatic_bounds},
and the loop models to be discussed below.

Now let $m$ be a positive integer, and let us decompose each weight $W(H)$
into $m$ nonnegative pieces, any way we like:
$W(H) = \sum_{\alpha=1}^m W_\alpha(H)$.
The first step of our generalized Chayes--Machta algorithm,
given a bond configuration $A$, is to choose, independently
for each connected component $H_i$, a ``color'' $\alpha \in \{1,\ldots,m\}$
with probabilities $W_\alpha(H_i)/W(H_i)$;
this color is then assigned to all the vertices of $H_i$.
The vertex set $V$ is thus partitioned as $V = \bigcup_{\alpha=1}^m V_\alpha$.
It is not hard to see that, conditioning on this decomposition,
the bond configuration is nothing other than a
generalized RC model with weights $\{ W_\alpha(H) \}$
on the induced subgraph $G[V_\alpha]$, independently for each $\alpha$.

We now have the right to update these generalized RC models
by any valid Monte Carlo algorithm.
One valid update is ``do nothing'';
this corresponds to the ``inactive'' colors
of Chayes and Machta \cite{Chayes-Machta}.
Of course, we must also include at least one nontrivial update!
The basic idea is to have at least one color
for which the weights $W_\alpha(H)$ are ``easy'' to simulate.
The original Chayes--Machta choice,
when $W(H) = q$ for all $H$,
is to take $W_\alpha(H) = 1$ for one or more colors $\alpha$
(the so-called ``active'' colors);
the corresponding model on $G[V_\alpha]$ is then independent bond percolation,
which can be trivially updated.
Since we must have $W_\alpha(H) \le W(H)$, this works whenever $q \ge 1$.

We hope that this explication/extension of the Chayes--Machta framework
will inspire other researchers to invent diverse algorithms
of Chayes--Machta type.
Here we would like to exhibit two such families of algorithms,
for simulating a general class of ``loop models'' on two-dimensional lattices.
More precisely, the models we have in mind should be called
{\em Eulerian-subgraph models}\/,
because the connected components are not necessarily loops;
we recall that a bond configuration $A$ is called Eulerian
if every vertex has even degree
(i.e., an even number of incident occupied bonds; zero is allowed).
So a generalized RC model is an Eulerian-subgraph model
if $W(H) = 0$ whenever $H$ is not Eulerian.
The simplest such model (the ``standard Eulerian-subgraph model'')
has
\be
   W(H)  \;=\;
   \cases{ n   & if $H$ is Eulerian \cr
           0   & otherwise \cr
         }
  \label{def.standard}
\ee
Another such model (the ``disjoint-loop model'') has
\be
   W(H)  \;=\;
   \cases{ n   & if $H$ is a cycle or an isolated vertex \cr
           0   & otherwise \cr
         }
  \label{def.disjoint-loop}
\ee
These two models are identical if the underlying graph $G$
has maximum degree $\le 3$ (e.g.\ the honeycomb lattice)
but are different otherwise.
More generally, we can consider the ``degree-weighted model''
\be
   W(H)  \;=\;  n \prod_{i \in V(H)}  \! t_{d_H(i)}
 \label{def.degree-weighted}
\ee
where $d_H(i)$ is the degree of the vertex $i$ in the graph $H$,
and $t_0, t_1, t_2, \ldots$ are nonnegative weights
satisfying $t_d = 0$ for all odd degrees $d$.

It is well known \cite{Domany_81} that
on any graph $G$ of maximum degree $\le 3$,
the model \reff{def.standard}/\reff{def.disjoint-loop}
arises for positive integer $n$
as a loop representation of an $n$-component spin model,
\be
   \widetilde{Z}
   \;=\; \mbox{Tr} \prod_{ij \in E} (1 + n x_{ij} \bsigma_i \cdot \bsigma_j)
   \;,
 \label{def_Z_spin}
\ee
where $\bsigma = (\sigma^1, \ldots, \sigma^n) \in \R^n$
and Tr denotes normalized integration with respect to any
{\em a priori}\/ measure $\< \,\cdot\, \>_0$ on $\R^n$ satisfying
$\< \sigma^{\alpha} \sigma^{\beta} \>_0 = n^{-1} \delta^{\alpha\beta}$
and $\< \sigma^{\alpha} \>_0 =
     \< \sigma^{\alpha} \sigma^{\beta} \sigma^{\gamma} \>_0 = 0$.
(In particular, uniform measure on the unit sphere is allowed,
 as are various ``face-cubic'' and ``corner-cubic'' measures \cite{Domany_81}.)
For $n \neq 1$, the Boltzmann weight \reff{def_Z_spin}
with spins on a sphere defines a non-standard $O(n)$ spin model,
which has positive weights only for $|x_{ij}| < 1/n$,
but it is nevertheless expected to belong to
the usual $O(n)$ universality class.

Likewise, on any graph $G$, the model \reff{def.standard}
[but not \reff{def.disjoint-loop}]
arises from the spin model \reff{def_Z_spin}
with a ``face-cubic'' {\em a priori}\/ measure,
i.e.\ $\bsigma$ is a unit vector $\pm {\bf e}_\alpha$ ($1 \le \alpha \le n$)
with probability $1/2n$.

Finally, let us observe that, on any {\em planar}\/ graph $G$,
there is a one-to-two correspondence between
Eulerian bond configurations $A$ on $G$
and Ising configurations $\{ \sigma \}$ on the dual graph $G^*$
(namely, $A$ is the Peierls contour for $\{ \sigma \}$).
Under this mapping, the bond model \reff{def.standard} on $G$ with $n=1$
is isomorphic to the Ising model on $G^*$
with couplings $J_e$ satisfying $v_e = e^{-2 J_e}$.
More generally, the model \reff{def.degree-weighted}
with $n=1$ and $t_d = a b^d$, i.e.
\be
   W(H)  \;=\;
   \cases{ a^{|V(H)|} b^{2|E(H)|}   & if $H$ is Eulerian \cr
           0                        & otherwise \cr
         }
  \label{def.pureexp}
\ee
gives an Ising model with $b^2 v_e = e^{-2 J_e}$.

With this observation, we can present two algorithms of Chayes--Machta type
for simulating Eulerian-subgraph models on a planar graph $G$.
In these algorithms, we keep at all stages
an Eulerian bond configuration $A$ on $G$
together with one of the corresponding Ising spin configurations
$\{\sigma\}$ on $G^*$.

The first algorithm applies to those models
in which there exist $a,b>0$ such that
$W(H) \ge a^{|V(H)|} b^{2|E(H)|}$ for all Eulerian $H$.
This includes in particular the model \reff{def.degree-weighted}
with $n \ge 1$ and $t_d \ge a b^d$.

{\em Algorithm 1:}\/ We use two colors:
an ``active'' color ($\alpha=1$) with $W_1(H)$ given by \reff{def.pureexp},
and an ``inactive'' color ($\alpha=2$) carrying the remaining weight.
The first step of the algorithm leads to a partitioning $V = V_1 \cup V_2$.
We now freeze all bonds having one or both vertices in $V_2$
in their current state, while leaving bonds within $V_1$ free to move.
The latter bonds are updated by applying one step of the
Swendsen--Wang algorithm to the correspondingly constrained
Ising model on $G^*$.
In detail, the algorithm proceeds as follows
(see \cite{fullpaper} for a proof of validity):

   1) Independently for each component $H_i$, color it $\alpha=1$
      with probability $W_1(H_i)/W(H_i)$, and $\alpha=2$ otherwise.

   2) On each edge of $G^*$ whose dual does not lie entirely within $V_1$,
      place a bond.
      On each edge $ij$ of $G^*$ whose dual $e$ lies entirely within $V_1$
      and for which $J_e \sigma_i \sigma_j > 0$,
      place a bond with probability $1 - e^{-2 |J_e|}$.
      (Here $J_e$ is defined by $b^2 v_e = e^{-2 J_e}$.)

   3) Form clusters on $G^*$ from sites connected by bonds.
      Independently on each cluster, flip the Ising spins with
      probability 1/2.
      (Note that a cluster typically contains spins of both signs.)

   4) The new bond configuration is the Peierls contour
      for the new Ising configuration.

This algorithm is easily extended to allowing more than one active color,
in case $W(H) \ge k a^{|V(H)|} b^{2|E(H)|}$ for some integer $k \ge 2$.
We freeze all bonds having one or both vertices in an inactive color
as well as all bonds connecting two distinct active colors.

Please note that Algorithm 1 is actually a {\em family}\/ of algorithms
corresponding to different choices of $a,b$.
Among the maximal allowed pairs $(a,b)$,
some choices may be more efficient than others.

Alternatively, one can use other choices of $W_1(H)$ and then update
the resulting bond model using either a local algorithm \cite{Karowski_83}
or a worm-type algorithm \cite{worm_algorithm}.

Our second algorithm applies to the standard Eulerian-subgraph model
\reff{def.standard} with $n \ge 1$.
Let us first observe, using the relation
cycles = bonds $-$ vertices + components,
that we can replace $n$ by $n^{c(H)}$ [$c(H) = $ cyclomatic number of $H$]
in \reff{def.standard} if we simultaneously replace
$v_{ij}$ by $x_{ij} = v_{ij}/n$.
Note also that the total number of {\em faces}\/
(including the exterior face) in a bond configuration $A$
equals its cyclomatic number plus 1.
Therefore, we can attribute a weight $n$ to each face.

{\em Algorithm 2:}\/ We again use two colors, but this time
we color the {\em faces}\/ of $A$
(which are clusters of vertices of the dual graph $G^*$).
In detail, the algorithm proceeds as follows
(see again \cite{fullpaper} for a proof of validity):

   1) Independently for each face of $A$, color it $\alpha=1$
      with probability $1/n$ and $\alpha=2$ with probability $1-1/n$.
      This decomposes the vertex set of $G^*$ as $V_1 \cup V_2$.

   2) On each edge of $G^*$ that does not lie entirely within $V_1$,
      place a bond.
      On each edge $ij$ of $G^*$ that lies entirely within $V_1$
      and for which $J_e \sigma_i \sigma_j > 0$,
      place a bond with probability $1 - e^{-2 |J_e|}$.
      (Here $e$ is the bond of $G$ dual to $ij$,
       and $J_e$ is defined by $x_e = e^{-2 J_e}$.)

     3,4)  As in Algorithm 1.

This algorithm is again easily extended to allowing more than one active color,
in case $n \ge 2$, using the same rule as in Algorithm 1.

\paragraph{Numerical results.}

We began by investigating the
loop model \reff{def.standard}/\reff{def.disjoint-loop}
on the honeycomb lattice,
for which Nienhuis \cite{Nienhuis_82+84}
and Baxter \cite{Baxter_86+87} found
the exact critical point when $-2 \le n \le 2$:
$x_c(n) = (2 + \sqrt{2-n})^{-1/2}$.
Nienhuis further observed \cite{Nienhuis_82+84,Nienhuis_91}
that the critical $O(n)$ model with $n \ge 0$ corresponds
to a {\em tricritical}\/ Potts model with $q = n^2$.
{}From
Coulomb-gas theory,
the leading and subleading thermal exponents
and leading magnetic exponent for the Potts model are
known to be
\cite{Nienhuis_82+84}
\be
   y_{t0} = \frac{3g-6}{g} \,,
   \;\;
   y_{t1}  = \frac{4g-16}{g} \,,
   \;\;
   y_{h0}  = \frac{(g+2)(g+6)}{8g}
 \label{critical_exponent}
\ee
where $g$ is the Coulomb-gas coupling and $q = 4 \cos^2 (\pi g/4)$;
here $g \in [2,4]$ for critical and $[4,6]$ for tricritical.
We also have a hull exponent \cite{Dup-Sal-Coniglio}
$y_H = (g+2)/g$.

We simulated the honeycomb-lattice loop model
\reff{def.standard}/\reff{def.disjoint-loop}
at $x = x_c(n)$ for $n=1.25$, 1.5, 1.75 and 2,
using Algorithm 1,
for 14 linear lattice sizes $L$ in the range $4 \leq L \leq 1024$.
Periodic boundary conditions were applied \cite{note_periodic}.
%
%
We measured the observables
$\scrn =$ number of occupied bonds,
$\scrs'_2 =$ sum of squares of loop lengths,
$\scrd_2 =$ sum of squares of face sizes,
and $\scrm =$ total Ising magnetization.
{}From these we obtained the specific-heat-like quantity
$C= L^{-d} (\langle \scrn^2 \rangle - \langle \scrn \rangle^2)$ 
and the Ising susceptibility $\chi_{\rm Is} = L^{-d} \langle \scrm^2 \rangle$
(here $d=2$).
%

Our numerical results for the static critical exponents
are shown in Table~\ref{table_1}.
We find empirically that
$\chi_{\rm Is} \propto L^{2y_{t0}-d}$,
$L^{-d} \< \scrn \> \propto {\rm const} + L^{y_{t1}-d}$,
$C \propto {\rm const} + L^{2y_{t1}-d}$,
$L^{-d} \< \scrd_2 \> \propto L^{2y_{h0}-d}$
and $L^{-d} \< \scrs'_2 \> \propto L^{2y_{H}-d}$.
(We shall give elsewhere \cite{Ding_06,fullpaper}
 theoretical arguments for several of these.)
The leading Potts exponents $y_{t0}$ and $y_{h0}$
are absent in the spin and loop observables of the $O(n)$ model,
but they can be seen in observables associated to the faces.
To our knowledge, this is the first observation of
$y_{t0}$, $y_{h0}$ and $y_H$
in $O(n)$ models
(see also \cite{Ding_06}).

\begin{table}
\begin{center}
\begin{tabular}{rr|llll}
  \multicolumn{1}{c}{$N$}        &
                                 &
  \multicolumn{1}{c}{$y_{t0}$}   &
  \multicolumn{1}{c}{$y_{t1}$}   &
  \multicolumn{1}{c}{$y_{h0}$}   &
  \multicolumn{1}{c}{$y_{H}$}    \\
\hline
  1.25   & Exact  & 1.83277   & 0.88740  & 1.93436    & 1.38908 \\
         & Num.   & 1.8327(1) & 0.887(2) & 1.9343(2)  & 1.3890(1) \\
  1.50   & Exact  & 1.78054   & 0.74811  & 1.91989    & 1.40649 \\
         & Num.   & 1.7805(1) & 0.749(5) & 1.9198(1)  & 1.4065(1) \\
  1.75   & Exact  & 1.70786   & 0.55428  & 1.90347    & 1.43071 \\
         & Num.   & 1.7080(1) & 0.55(1)  & 1.9035(1)  & 1.4307(1) \\
  2.00   & Exact  & 1.5       & 0        & 1.875      & 1.5 \\
         & Num.   & 1.5000(1) & $-0.01(2)$   & 1.8750(1)  & 1.5001(1)
\end{tabular}
\end{center}
\vspace*{-5mm}
\caption{
   A comparison of the critical exponents determined by
   Monte Carlo simulations and those predicted 
   by \reff{critical_exponent} {\em et seq.}\/
}
\label{table_1}
\end{table}

\begin{figure}[t]
\begin{center}
\leavevmode
\includegraphics[scale=0.6]{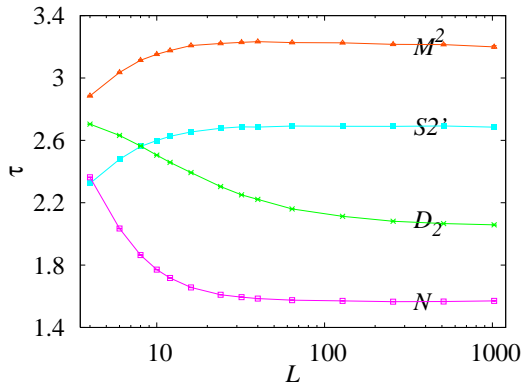}
\end{center}
\vspace*{-6mm}
\caption{
   Integrated autocorrelation times $\tau_{\rm int}$
   for the $O(2)$ loop model on the honeycomb lattice
   versus lattice size $L$.
}
\label{fig_tau2}
\end{figure}

We have also determined the integrated autocorrelation time $\tau_{\rm int}$
for the observables $\scrn$, $\scrs'_2$, $\scrd_2$ and $\scrm^2$.
For $1 < n \le 2$, critical slowing-down is either entirely absent
or very nearly absent.
The $\tau_{\rm int}$ data for $n=2$ are shown in Fig.~\ref{fig_tau2},
and they strongly suggest that the dynamic critical exponent $z$
is zero for $n=2$.

\begin{figure}[t]
\begin{center}
\leavevmode
\includegraphics[scale=0.61]{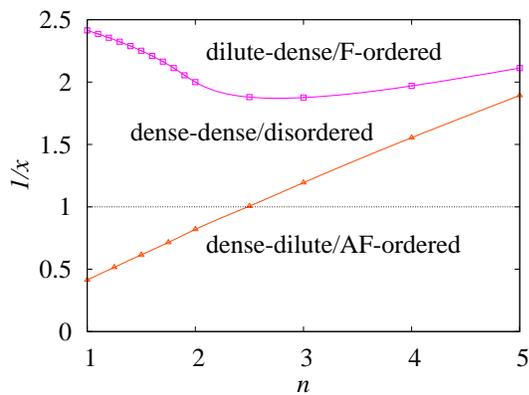}
\end{center}
\vspace*{-6mm}
\caption{
   Location of the ferro- and antiferromagnetic phase transitions
   of the standard Eulerian-subgraph model \reff{def.standard}
   on the square lattice, as a function of $n$.
}
\label{fig_fcsq}
\end{figure}

We next simulated
model \reff{def.standard}
on the square lattice, using Algorithm 2,
for lattices of linear size $4 \le L \le 512$
with periodic boundary conditions \cite{note_periodic}.
The phase diagram of this model is largely unexplored,
especially when $n > 2$.
Phase transitions were located by
analyzing the data for several Binder-type ratios.
Two distinct phase transitions were found, 
which correspond to the ferromagnetic and antiferromagnetic
transitions in terms of the dual Ising spins
(see Fig.~\ref{fig_fcsq}).
The three phases are, respectively,
dilute--dense, dense--dense and dense--dilute,
where A--B means that the bond configuration is of type A
and its complement is of type B (here ``dense'' = ``percolating'').
The locations of the ferromagnetic critical point for $n \le 2$
agree accurately with those given in \cite{Guo_06_1}.
Our data show that the lines of phase transitions extend to $n>2$.
We shall discuss elsewhere \cite{fullpaper}
the nature of this new phase transition.
The critical slowing-down is essentially absent for $1 \le n \le 2$,
but is strong for $n > 2$.
%

We also applied Algorithm 1 to model \reff{def.degree-weighted}
on the square lattice.
The results, along with a conjectured RG flow in the $(n,v,t_4)$
space, will appear elsewhere \cite{fullpaper}.


\begin{acknowledgments}
This work was supported
by
NSF grants PHY--0116590/PHY--0424082
and NSFC project 10675021.
\end{acknowledgments}

\vfill


\begin{thebibliography}{99}
\bibitem{note_loop_rep}  The loop representation works in its simplest form
   only for lattices of maximum degree $\le 3$
   (e.g.\ the honeycomb lattice)
   and for a special form of the Boltzmann weight
   ($1 + x \bsigma_i \cdot \bsigma_j$ instead of
    $e^{\beta \bsigma_i \cdot \bsigma_j}$).
\end{thebibliography}
\end{document}